# Specific heat of $Nb_3Sn$: The case for a second energy gap

V. Guritanu[1], W. Goldacker[2], F. Bouquet[1,†], Y. Wang[1], R. Lortz[1], G. Goll[3], and A. Junod[1,‡].

[1] Department of Condensed Matter Physics, University of Geneva, CH-1211 Geneva 4 (Switzerland).

[2] Forschungszentrum Karlsruhe, Institut für Technische Physik, P.O. Box 3640, D-76021 Karlsruhe (Germany).

[3] Physikalisches Institut, Universität Karlsruhe, D-76128 Karlsruhe (Germany).

## Abstract

We present new specific heat data for $Nb_3Sn$, a well-known technically applied superconductor with a critical temperature $T_c \cong 18$ K, in the temperature range from 1.2 to 200 K in zero magnetic field, and from 1.5 to 22 K in fields $H \leq 16$ T. The particularly dense and homogeneous polycrystalline sample used for this study is characterized in detail. We determine the bulk upper critical field $H_{c2}(T)$ from specific heat data, and the Sommerfeld constant $\gamma$ from the entropy $S(T)$. We investigate in detail a low-temperature anomaly already noticed in previous investigations in zero field, and find that this feature can be quantitatively ascribed to the presence of a second superconducting gap $2\Delta_S(0) \cong 0.8 k_B T_c$, in addition to the main one $2\Delta_L(0) \cong 4.9 k_B T_c$. The signature of this minor gap, which affects 7.5% of the electronic density-of-states, vanishes in high fields.

## Keywords

$Nb_3Sn$, specific heat, superconducting gap, upper critical field

## PACS numbers

74.70.Ad (A15 materials), 74.25.Bt (thermodynamic properties), 74.20.De (phenomenological theories), 74.25.Op (critical fields), 74.25.Jb (EDOS of superconductors)

† Present address: Laboratoire de Physique des Solides, Université de Paris Sud XI, F-91405 Orsay Cedex (France).

‡ Corresponding author. e-mail: alain.junod@physics.unige.ch

**Introduction**

The unexpected discovery of superconductivity at 39 K in $MgB_2$, a classical non-oxide intermetallic compound, has stimulated a great deal of interest. A particularly interesting feature of $MgB_2$ is the existence of two sets of electronic bands crossing the Fermi level, which give rise to distinct superconducting gaps.[1] The presence of two gaps (more precisely, two groups of gaps) was assessed by various techniques probing the surface as well as the bulk of the sample (see Ref. [2] for a review). The two-gap feature is particularly evident in $MgB_2$ because the density of states at the Fermi level is almost equally shared between the two sets of bands, and because the gaps widths, being in a ratio ~3:1, are sufficiently different to be easily resolved. One may wonder if two-gap superconductivity only occurs as an exception in nature, being limited to $MgB_2$, Nb-doped $SrTiO_3$ [3] and possibly some *s-d* elements,[4, 5] or if it exists more generally, but has not been paid sufficient attention in the past. One of the conditions for multigap superconductivity to occur is that more than one band should cross the Fermi level, a prerequisite that is commonly satisfied e.g. in *s-d* metals. A second condition is weak interband scattering.[6] The latter is seldom met. The different dimensionality characterizing the $\sigma$ and $\pi$ bands in $MgB_2$ helps to satisfy this requirement.[7]

The specific heat of some superconductors differs markedly from the BCS behavior at low temperature, indicating that they may be candidates in the search for additional examples of multigap superconductivity. One of them is $Nb_3Sn$, a material which like NbTi is in technological use, and in which the structural, electronic, magnetic, elastic, vibrational, and superconducting properties have been widely documented.[8-11] Several studies have shown that the specific heat of $Nb_3Sn$ does not vanish exponentially at low temperature below $\sim T_c / 4$; these studies were generally limited to zero field or were not sufficiently detailed, so that no definitive conclusions could be drawn. In this paper, we reinvestigate the specific heat in the temperature range from 1.2 K to well over $T_c$, in several magnetic fields up to 16 T, in order to decide if the origin of this anomaly lies in the lattice vibrations or in the electronic spectrum. We find that this anomaly is field-dependent, which excludes phonons as a possible cause. Furthermore, we find that the two-gap model recently advocated by Bouquet *et al.* [12] accurately fits our data. Finally, very recent point-contact spectroscopy experiments performed on the same sample also detect a feature on the same energy scale as the smaller gap.[13] We conclude that $Nb_3Sn$ is a new example of two-gap superconductivity.

**Previous specific heat studies**

We first review earlier representative work in order to establish that the anomaly we discuss is a distinctive and general feature of $Nb_3Sn$ rather than the property of a particular sample.

The specific heat data in zero field of Vieland and Wicklund [14] follow the law $C/T = \gamma_0 + \beta_3 T^2$ between 1.1 K and 5 K, with $\gamma_0 \cong 0.8$ mJ/(K$^2$gat) (one gat is ¼ of a mol, i.e. 99.5 g of $Nb_3Sn$). Ideally, in a fully gapped superconductor, the parameter $\gamma_0$ should be zero. The term $\beta_3 T^2$ represents the lattice contribution. Vieland and Wicklund noted that this behavior is consistent either with one-dimensional modes in the vibrational spectrum, or with the two-gap model of Suhl *et al.* [6] in the limit $k_B T >> \Delta_S(0)$, where $\Delta_S(0)$ is the smaller gap attributed to the *s*-band; in this regime the *s*-band electrons make a normal electron-like contribution to the specific heat. However the decreasing specific heat in the fully gapped regime $k_B T << \Delta_S(0)$ was not observed, so that this interpretation remained speculative.

One of us (A. J.) investigated the specific heat of several $Nb_3Sn$ polycrystals with various $T_c$s in the same temperature range.[15] For two nearly stoichiometric samples with $T_c \cong 18$ K, it was found that $C/T$, although similar to Vieland's data above 2.2 K, showed a downturn as $T \to 0$, which was inconsistent with a constant value of the $\gamma_0$ parameter. It was suggested that this anomaly was related to the softening of the $[\varsigma\varsigma 0]T_1$ phonon mode, which had been observed in neutron scattering experiments.[16]

Stewart *et al.* also found an anomalous specific heat for $1.4 \leq T \leq 4.5$ K in a polycrystalline sample which, according to the lattice constant $a = 5.290$ Å, was probably slightly Sn-deficient.[17] Using the analytic form $C/T = \gamma_0 + \beta_3 T^2$, they obtained $\gamma_0 \cong 0.25$ mJ/(K$^2$gat). In a later work, the value $\gamma_0 \cong 0.5$ mJ/(K$^2$gat) was given for a single crystal, using data for $2.0 \leq T \leq 4.2$ K.[18] Stewart *et al.* listed several possible explanations for $\gamma_0$ being non-zero, including two-gap superconductivity and a phonon anomaly, and rejected the possible effect of impurities.

Among other measurements of the specific heat of $Nb_3Sn$ available in the literature, the precise data of Khlopkin in fields up to 19 T should be mentioned; however the temperature range below 4.5 K was not investigated.[19]

**Sample**

The polycrystalline sample used for this study was synthesized at the Forschungszentrum Karlsruhe by hot isostatic pressing (HIP). Powders of Nb (99.9% purity) and Sn (99.5%) were mixed in a nominal composition $Nb_{0.746}Sn_{0.254}$ and reacted at 1100°C for 24 h under a pressure $p(Ar) = 100$ bar. The excess tin provides a liquid phase during the whole process and helps to form dense samples; the resulting macroscopic density is greater than 98.3% of the X-ray density of 8.91 g/cm$^3$ calculated with the assumption of a perfect crystal lattice. The homogeneity, as indicated by the sharpness of the superconducting transition (see below), is high even for large samples. Details and extensive characterization are given in Ref. [20].

Powder diffraction patterns were obtained using a Huber G645 diffractometer in the Guinier geometry, and were refined by the Rietveld method using the FULLPROF program.[21] Above 50 K, the patterns only show the lines of the cubic A15 phase, space group $Pm3n$ (Fig. 1). The lattice parameter at room temperature, $a = 5.293$ Å, is consistent with a slight excess of Sn, indicating that the final composition is nearly identical to the nominal one, $Nb_{0.746}Sn_{0.254}$.[20] Near 40 K, the sample undergoes a structural martensitic transformation (Fig. 1, inset). The tetragonal distortion, $c/a - 1 = -0.0026$ at 10 K, is approximately half that reported for a single crystal in the literature;[22] it is estimated that only 30 to 40% of the sample volume transforms.[20] We recall that the martensitic transformation is highly sensitive to the Sn concentration, and only occurs in the range of compositions between approximately $Nb_{0.747}Sn_{0.253}$ and $Nb_{0.755}Sn_{0.245}$.[20, 23] There is a small non-transforming phase regime between $Nb_{0.747}Sn_{0.253}$ and the phase boundary located at approximately $Nb_{0.744}Sn_{0.256}$. The composition of the sample reported in this paper was chosen to be in the tin-rich non-transforming regime. Systematic studies have shown that the tetragonal distortion does not lower the superconducting transition temperature by more than a fraction of a degree;[24] neither does it significantly affect the density of states at the Fermi level.[25, 26]

The superconducting transition temperature was determined by three methods: *ac*

susceptibility in an alternating field of 10 $\mu$T at 80 Hz, *dc* magnetization measured in a SQUID magnetometer at 0.5 and 5 mT, and the specific heat jump in zero field, which, in this order, are increasingly representative of the bulk. The transition width is $\Delta T_c = 10$ mK, 10%-90%, according to *ac* susceptibility (Fig. 2a, inset), and 80 mK, 10%-90%, according to the specific heat jump. The *dc* magnetization is consistent with full diamagnetism (Fig. 2b, inset). The midpoint of the specific heat jump is located at 17.75 K, ~0.3 K below the midpoint of the *ac* susceptibility step. The latter, which is sensitive to shielding, may indicate a different tetragonal distortion at the surface due to local strain. Indications for this scenario can be found in the rather large anomaly (compared to our specific heat data) observed in the thermal expansivity at the martensitic transition,[20] which is directly linked to a strong strain/stress dependence of this structural transition.

A bar-shaped sample with a length of 5.8 mm and a cross-section of 0.56 mm$^2$ was cut from the main sample and used for standard 4-wire resistance measurements. The resistivity just above the superconducting transition is $\rho(T_c) = 13.0 \pm 0.2$ $\mu\Omega$ cm, which is about half the best values reported for polycrystals,[20, 23] and 4-20% larger than those obtained for vapor-deposited films;[27] $\rho(300 \text{ K})/\rho(T_c) = 5.4$.

**Calorimetry and data analysis**

For measurements from 15 to 200 K in zero field, a 0.30 g piece was cut from the main sample, and measured in an adiabatic, continuous-heating calorimeter.[28] Results for $C(T)$ are shown in Fig. 2a. No singularity was observed at the martensitic transformation temperature to within the limits of the experimental scatter of ~0.02%; only a small change in the slope was barely detectable. In the upper temperature range, the specific heat is dominated by the lattice contribution. The effective Debye temperature is defined as the value $\Theta_D(T)$ such that the calculated Debye specific heat $C_D(\Theta_D/T)$ is equal to the measured lattice specific heat per gram-atom $C_{ph}(T)$ at a given temperature $T$. The full Debye specific heat function $f_D(\Theta_D/T) = C_D/3R$ is calculated numerically. The "instantaneous" Debye temperature is then obtained from $\Theta_D(T) = T f_D^{-1}(C_{ph}(T)/3R)$. The electronic contribution $C_e$ must be subtracted first. Up to $T \approx 0.2\Theta_D$, we use the low-temperature Sommerfeld contribution $C_e = \gamma T$ as determined below. Above $T \approx 0.3\Theta_D$, i.e. in the upper half of the temperature range of the present measurements, the electron-phonon renormalization contribution is expected to vanish, therefore we subtract an electronic specific heat reduced by a factor of $1 + \lambda \approx 3$.[29, 30] In practice, the uncertainty on the electronic contribution remains relatively unimportant over a broad temperature range. Anharmonic corrections are neglected. Within these assumptions, we find that the effective Debye temperature increases monotonously from $\Theta_D(0) = 234$ K (see below) and $\Theta_D(T = 200\text{K}) \cong 350$ K.

For measurements below 22 K as a function of the magnetic field, we cut a 21 mg piece from the previous sample and used a relaxation calorimeter.[28] A particular feature of our technique is that each thermal relaxation provides 10-100 data points, acquired during both the heating and cooling periods. The field is always applied or changed above $T_c$. The zero-field curve $C(T)$ shows a sharp jump $\Delta C / T_c = 34\,\text{mJ/(K}^2\text{gat)}$ at the superconducting transition (Fig 2b; also see Fig. 6b below), comparable with the literature data 31 mJ/(K$^2$gat),[31] 32 mJ/(K$^2$gat),[15] 31 mJ/(K$^2$gat),[18] and 28 mJ/(K$^2$gat).[19]

The highest field available in our laboratory, 16 T, can suppress superconductivity in Nb$_3$Sn only down to ~10 K. Owing to the temperature dependence of the effective Debye temperature, the separation of the electronic and lattice heat capacities requires some care. We assume that for $T < 22$ K, the normal-state specific heat obeys the usual form

$$C_n(T) = \gamma T + \beta_3 T^3 + \beta_5 T^5 ,$$

where $\gamma T$ is the Sommerfeld electronic contribution (not to be confused with the anomalous residual term $\gamma_0 T$), and $\beta_3 T^3 + \beta_5 T^5$ are the first terms of the low-temperature expansion of the lattice specific heat. The unconstrained extrapolation of the normal-state specific heat $C_n / T$ versus $T^2$ from ~10 K to zero would introduce a large uncertainty, in particular on $\gamma$ because $\gamma T$ provides a minor contribution above ~10 K. Therefore we do not determine the normal-state curve $C_n(T)$ by fitting the measured specific heat $C_s(T)$ above $T_c(H)$; we rather determine the normal-state entropy $S_n(T)$ by fitting the measured entropy $S_s(T)$ above $T_c(H)$. This automatically satisfies the constraint $S_n(T_c) = S_s(T_c)$. In addition, phonons contribute ~3 times less in the low-temperature expansion of the normal-state entropy:

$$S_n(T) = \gamma T + \frac{1}{3}\beta_3 T^3 + \frac{1}{5}\beta_5 T^5 ,$$

Evidently, $C_n(T)$ is determined once $S_n(T)$ is known. Experimentally, the entropy is obtained by numerical integration of the data, taking into account the third law of thermodynamics:

$$S(T) = \int_0^T \frac{C}{T'} dT'$$

Missing data between 0 and ~1.2 K are extrapolated using the empirical expression $C_s(T \ll T_c)/T = \gamma(H) + aT^n$ fitted between ~1.2 and 4 K. As shown in Fig. 3, the extrapolation of $S_n/T$ versus $T^2$ from $T_c(H)$ to $T = 0$ is well defined. It yields $\gamma = 13.7$ mJ/(K$^2$gat), using a global fit of all $S_n(T)$ curves between $T_c(H)$ and 22 K. Literature data show some scatter as they are usually based on an extrapolation of $C/T$: 13.1 mJ/(K$^2$gat),[31] 11.2 to 13.5 mJ/(K$^2$gat),[15] 8.3 mJ/(K$^2$gat),[18] and 11 mJ/(K$^2$gat).[19] The same fit determines the lattice coefficients $\beta_3 = 0.152$ mJ/(K$^4$gat) and $\beta_5 = -0.077$ $\mu$J/(K$^6$gat). The initial Debye temperature obtained from

$$\Theta_D(0) = \left(\frac{12R\pi^4}{5\beta_3}\right)^{1/3} ,$$

where $R$ is the ideal gas constant and $\beta_3$ refers to one gram-atom, is $\Theta_D(0) = 234$ K. Comparable data are found in the literature: 228 K,[31] 225 to 238 K,[15] 230 K,[18] and 232 K.[19] This determination of the lattice specific heat allows us to isolate the electronic contribution, assuming as usual that superconductivity does not measurably affect the lattice component:

$$C_e(T) = C(T) - \beta_3 T^3 - \beta_5 T^5 .$$

The condensation energy and the thermodynamic critical field are obtained by integration of experimental data, using the cell volume at 10 K (Table I). With a maximum of +0.026 at $t \equiv T/T_c = 0.62$, the deviation function $D(t) \equiv h - (1-t^2)$, where $h \equiv H_c(T)/H_c(0)$, is comparable with that of Pb, and characteristic of strong coupling (Fig. 5, inset).[32, 33]

**Electronic specific heat in zero field**

The electronic specific heat in the superconducting state normalized to that in the normal state, $C_{es}/\gamma T$, has been tabulated by Mühlschlegel for an isotropic, single-band BCS superconductor.[34] This is represented by the long-dashed line in Figs 6a and 6b below; note that $C_{es}/\gamma T$ becomes negligible on the scale of these plots below $\sim 0.15T_c$. The data for Nb$_3$Sn differ significantly from this ideal behavior. In particular, the data below $\sim 0.15T_c$ exceed the BCS values by several orders of magnitude (see Fig. 6a below). This anomaly, which cannot be described by a simple constant $\gamma_0$, is the key point of the present study.

Near $T_c$, one also notices a large deviation with respect to the BCS model. The reduced specific heat jump $\Delta C(T_c)/\gamma T_c \cong 2.5$ definitely exceeds the BCS value 1.426. However, this difference is well understood in the framework of strong-coupling superconductivity, and merely means that the condition $T_c/\overline{\omega} << 1$, where $\overline{\omega}$ is an average phonon frequency, is not fulfilled.[33, 35] Using the empirical model of Padamsee *et al.*,[36] we find that the large jump at $T_c$ can be accounted for by assuming a gap $2\Delta_L(0) \cong 4.7k_BT_c$ rather than the BCS value $2\Delta(0) = 3.53k_BT_c$ (we note that this tentative fit exceeds the data for $0.45 < T/T_c < 0.9$, unlike the improved model introduced later). Owing to the larger gap, strong-coupling corrections imply an even faster decrease of the electronic specific heat as $T \to 0$ compared to the BCS model (see Fig. 6a below, short-dashed line); therefore strong coupling does not help to explain the anomaly at $T \to 0$.

**Electronic specific heat in a magnetic field**

We have measured the specific heat in fields of 2, 4, 7, 10, 13, and 16 T from 1.5 to 22 K. These measurements were later supplemented by low-field data at 0.2, 0.5, and 1 T from 1.5 to 5 K. The first data set is shown in Fig. 4. At each field, a specific heat jump marks the superconducting transition $T_c(H)$. The jump in zero field is larger than the continuation of the jumps at the transition from the normal state to the mixed state, as expected from Maki's theory,[37] and already documented experimentally by Khlopkin.[19] It is crucial to note that the zero-field anomaly at $T^2 < 20$ K$^2$ disappears at high fields (Fig. 4, insets). This rules out phonons as a possible cause of the anomaly.

Although the following point is not central for our discussion of the low-temperature anomaly, we use the present specific heat data to better define the bulk value of the upper critical field of Nb$_3$Sn at $T \rightarrow 0$. In the first method, we rely essentially on the initial slope of $H_{c2}(T)$, and fit the WHH formula in the dirty limit [38] to $T_c(H)$ as determined from the midpoint of the specific heat jump for $0 \leq H \leq 4$ T. The result is $H_{c2}(0) = 25$ T. However, the data at 10, 13 and 16 T noticeably exceed the WHH curve (Fig. 5), a deviation that generally occurs not only in strong-coupling superconductors, but also in anisotropic or two-band superconductors.[39, 40] Both $H_{c2}(0)$ and the deviations are in perfect agreement with the work of Khlopkin.[19] We alternatively determined the transition temperature in each field using the local entropy balance, $T_c(H)$ being then given by the intersection of the quasi-linear sections of the entropy curves in Fig. 3 extrapolated from above and from below into the transition region. The latter determinations are shown as an error bar on the low temperature side of the ● symbols in the phase diagram (Fig. 5). The difference with the "midpoint" is at most of the order of the radius of the symbols, so that the deviation with respect to the WHH curve appears to be a robust feature. In the second method, we look at the increase of the mixed-state electronic coefficient $\gamma_m(H) \equiv \lim_{T \rightarrow 0}[C(T,H)/T]$. This would require an extrapolation to $T = 0$; we show in the inset Fig. 3 the values of $C_e(T,H)/T$ at $T = 1$ K (see also Fig. 4, right inset). $\gamma_m(H)$ increases almost linearly as a function of the field, and is bound to reach the Sommerfeld constant $\gamma$ at $H_{c2}(0)$.[41] The intersection of the $\gamma_m(H)$ line with the $\gamma$ limit defines $H_{c2}(0) = 25$ T, in agreement with the previous method, without having to assume any functional shape for $H_{c2}(T)$. Resistively determined transitions for Nb$_3$Sn wires in static fields up to 23 T have yielded $H_{c2}(0) = 24.5$ T.[42]

We now concentrate on the anomaly in the specific heat at $T \rightarrow 0$ which is visible at the origin of Fig. 4. Most of the "wiggle" below $T^2 \approx 40$ K$^2$ vanishes between 2 and 4 T (see also the entropy in Fig. 3). The right inset of Fig. 4 shows an expanded view below 4.5 K, including additional data at low fields. This plot reveals that the residual anomalous negative curvature below $T^2 \approx 20$ K$^2$ persists up to ~7 T. At higher fields, the electronic specific heat in the mixed state follows the usual law

$$C_{es}(T,H)/T = \gamma_m(H) + \beta_m(H)T^2, \quad T << T_c(H)$$

with $\gamma_m(H) \cong \gamma H / H_{c2}(0)$; $\beta_m(H)T^2$ is a second-order term in the development of the mixed-state electronic specific heat.[37] Therefore the present data show a smooth crossover from an anomalous behavior at low field (< 2 T) to a standard behavior at high field (> 8 T). We shall refer to the intermediate field scale as the crossover field $H_{c2,s} \approx 5\pm3$ T, keeping in mind that its value depends on the selected criterion.

Several arguments show that the low-temperature anomaly cannot be explained by the superconducting transition of an impurity phase distributed between 2 and 6 K. The magnetization is perfectly flat below $T_c$, both in the field-cooled and zero-field cooled mode (inset of Fig. 2b). The critical field of a Sn-rich impurity would be inconsistent with the observed crossover field, since $H_c(0) \cong 0.03$ T for pure Sn. On the Nb-rich side, A15-type Nb$_{1-x}$Sn$_x$ can be excluded because of its minimum critical temperature of ~7 K. Finally X-ray patterns do not show any extra line, which sets a limit of ~3% for the concentration of a hypothetical second phase. In order to account for the amplitude of the anomaly with such a small concentration, the $\gamma$-value of the low-$T_c$ phase would have to be two to three times larger than that of Nb$_3$Sn itself. The latter is already unusually large. This is very unlikely.

**Discussion**

The specific heat data presented here display an anomaly in the specific heat below ~0.25 $T_c$ which disappears in the vicinity of 5±3 T. This behavior is reminiscent of the excess specific heat observed below ~0.6 $T_c$ in MgB$_2$ (inset of Fig. 6a), and which vanishes above ~0.5 T.[43] The maximum excess $C/T$ with respect to the BCS curve is ~$\gamma/2$ in MgB$_2$, compared with ~$\gamma/13$ for Nb$_3$Sn (Fig. 6a). This correspondence suggests that we may quantitatively analyze the zero-field data for Nb$_3$Sn using the empirical two-gap model of Bouquet *et al.* developed and successfully applied to MgB$_2$.[12] The fitted two-gap curve is shown together with the data at low temperature in Fig. 6a, and up to $T_c$ in Fig. 6b; the three parameters of the fit, i.e. two gaps and the DOS fraction, are listed in Table II. The two-gap model not only correctly represents the low temperature anomaly (except for a few points below $T_c/10$), allowing the width of the smaller gap $2\Delta_L(0)_c \cong 0.8 k_B T_c$ to be determined, but also significantly improves the fit in the intermediate temperature range $0.45 < T/T_c < 0.9$ as compared to the single-gap model. Note that the value of the larger gap has increased somewhat, $2\Delta_L(0)/k_B T_c \cong 4.9$ rather than 4.7. The larger gap combines with the reduced weight to match the same jump height, but the added degree of freedom in the two-gap model allows the change of slope $d(C/T)/d(T^2)$ at $T_c$ to be more accurately fitted than with the single gap model. Therefore the presence of a second gap in Nb$_3$Sn with parameters as given in Table II is consistent with the present data at all temperatures. The smaller gap $\Delta_S$ opens on a band (or set of bands) that represent 7.5% of the total density of states (more precisely, the partial Sommerfeld constant is $\gamma_S \cong 0.075\gamma$), and its width $\Delta_S(0)$ is about six times smaller than that of the main gap $\Delta_L(0)$. The contribution of the minor gap to the specific heat is reminiscent of that of a semiconducting gap.[2] The partial specific heat should initially increase exponentially for $k_B T << \Delta_S$; indications of such an increase are observed at the lowest temperatures of our experiment. When $k_B T > \Delta_S$ the behavior of the electrons in the small gapped band will essentially be identical to that of normal electrons $C_{es,S} \cong \gamma_S T$. At still higher temperature, the exponential increase due to the main gap takes over (Fig. 6a).

The BCS plot of the logarithm of the specific heat versus the inverse temperature gives another way to visualize the smaller gap (inset of Fig. 6b). Generally speaking, in the range $2.5 < T_c/T < 3.3$, the slope of $\ln(C_{es}/\gamma T_c)$ versus $T_c/T$ would appear to equal $-(1.44/1.76)\Delta(0)/k_B T_c$ to within 10% for a wide range of materials and model spectra.[44] In Nb$_3$Sn, one observes a crossover from a slope higher than BCS at high temperature, to a slope $-0.2$, much smaller than BCS, at low temperature. Again, this reflects one of the gaps being larger, and the other one smaller than the BCS gap, as required by the theory of two-band superconductivity.[45] However, because of large strong-coupling corrections, the gap parameters given in Table II do not satisfy the BCS sum rule of Combescot.[45] Note that the value $2\Delta_L / k_B T_c = 4.9$ obtained in this work somewhat exceeds those obtained from tunneling measurements on stoichiometric samples, which fall in the range 4.1 to 4.5.[46-48] Surface values may differ from bulk values; however it would be interesting to include two gaps in the inversion algorithm of the tunneling spectra to see how it would affect the results.[30]

The effect of the field can only be discussed at a qualitative level. In analogy with MgB$_2$, we define two values of the coherence length which are relevant for the carriers of the bands labeled $S$ and $L$, respectively: $\xi_S \sim \hbar v_{FS} / \pi \Delta_S$ and $\xi_L \sim \hbar v_{FL} / \pi \Delta_L$, where $v_{FS}$ and $v_{FL}$ are the Fermi velocities. Formally, they correspond to two critical fields $H_{c2,S}(0) \sim \Phi_0 \Delta_S^2(0) / \hbar^2 v_{FS}^2$ and $H_{c2,L}(0) \sim \Phi_0 \Delta_L^2(0) / \hbar^2 v_{FL}^2$. The first of these is a "crossover field" at which $S$-band vortices overlap, $S$-band electrons start to contribute as normal carriers, and the structure in $C/T$ associated with the smaller gap levels off. Unlike MgB$_2$, this crossover does not appear in the plot of $\gamma_m(H)$ (inset of Fig. 3). We attribute this to the smallness of the DOS of the $S$ band, and to the fact that $\gamma_m(H)$ is estimated at $T = 1$ K rather than $T = 0$, which adds a compensating positive curvature due to excited pairs (note that data taken at $T \ll 1$ K would not improve the determination, as they would be dominated by the hyperfine specific heat of Nb nuclei). The second field is the true upper critical field, at which $L$-band vortices overlap and at the same time superconductivity is suppressed.[49] Estimating $H_{c2,S}(0)$ from the field beyond which the anomalous low-temperature wiggle disappears in Fig. 4, we find $3 \le H_{c2,L} / H_{c2,S} \le 12$. This ratio does not have to scale with the gap ratio since the Fermi velocity also plays a role. Together with $\Delta_L(0) / \Delta_S(0) \cong 6$, we obtain $0.3 \le v_{FS} / v_{FL} \le 0.6$, i.e. the minor gap would be associated with slow carriers.

This result is unexpected. The identification of the band or group of bands which are associated with the smaller gap is uncertain. Five or six bands cross the Fermi level in the tetragonal or cubic phase of Nb$_3$Sn, respectively. The band structure is sensitive to structural details such as the dimerization of Nb chains (corresponding to a $\Gamma_{12}$ optical phonon) and, to a much lesser extent, the tetragonal distortion.[25, 26] The splitting of the $\Gamma_{12}$ doublet hardly affects the DOS at the Fermi level.[25, 26] Taking the Fermi surface sheets obtained by the linear muffin-tin orbital method, which are supported by two-dimensional angular correlation of positron annihilation radiation in the cubic state,[50] we can exclude some sheets because their contribution to the DOS is too large. This is the case for bands #5 and #6 in the numbering adopted by Ref. [50], #5 containing the Fermi surface of a very flat band due to the Nb 4d electrons, assumed to be responsible for the high $T_c$. The remaining candidates are mostly empty "jungle-gym" structures of holes. According to the decomposition of the total DOS at the Fermi level into symmetry-projected components given by Mattheiss and Weber for the cubic phase,[51] 83.4% of the bare DOS originates from the Nb $d$-bands, in particular $d(\sigma)$ and $d(\pi)$, the remaining part being essentially due to the Nb $p$-bands. Taking into account the large renormalization $1+\lambda \approx 3$ for $d$ states,[30] and assuming $1+\lambda \approx 1.2$ for the other ones, the fraction of the renormalized DOS which originates from the Nb $d$-bands becomes $\approx 92.6\%$. This is quantitatively consistent with the experimental ratio $1-x = 0.925$ (Table II), and gives support to a scenario in which the superconducting coupling originates from the Nb $d$-bands, while a minor gap or several small gaps are induced in the $s$- and $p$-bands by interband scattering or Cooper pair tunnelling (for a review see Ref. [52]).

Is the second gap a specific characteristic of "high" temperature A15 superconductors? The answer is probably no. We have inspected the low-temperature specific heat of three nearly stoichiometric samples of $V_3Si$ measured down to 1.35 K in our laboratory, and see no comparable anomaly. The data are strictly linear in the $C/T$ versus $T^2$ plot below 5 K$^2$, and point to $\gamma_0 = 0.04$, 0.02 and 0.00$\pm$0.001 mJ/(K$^2$gat), for a polycrystal and two single crystals, respectively. The single crystal data of Ref. [53] follow a similar behavior with $\gamma_0 = 0.05$ mJ/(K$^2$gat) down to 0.3 K, without any trace of a downturn. As for the other A15 compounds, some samples, in particular the non-stoichiometric ones, tend to show a larger value of $\gamma_0$ that may be simply attributed to incomplete superconductivity or second phases.

Finally we would like to comment on the possible anisotropy of the main gap $\Delta_L$. It was reported in the early tunneling measurements of Hoffstein and Cohen that $2\Delta(0)/k_BT_c = 2.8$, 2.1, and 1.0 along the [100], [110], and [111] directions, respectively.[54] The minimum gap might be consistent with the anomaly we observe in the specific heat; the other values are not. However, more recent high quality data on superconductor / insulator / superconductor junctions exclude gap anisotropy and give $2\Delta(0)/k_BT_c = 4.1$.[30, 48] Furthermore the measured specific heat is not compatible with a standard anisotropy for which $\Delta(\vec{k})$ is described by an ellipsoid.

**Conclusion**

The new specific-heat measurements presented in this work on a particularly homogeneous polycrystalline sample give more details on a low-temperature anomaly that has previously been reported in $Nb_3Sn$, and in particular on its behavior in a magnetic field. The observed features are consistent with the presence of a second gap about 6 times smaller than the main one. This minor gap opens in a band or group of bands which are responsible for ~7.5% of the total renormalized DOS, most probably the Nb and Sn $s$ and $p$ bands. Its signature vanishes at high fields.

One may wonder if this interpretation of the low-temperature specific heat of $Nb_3Sn$ is unique. From the variation of the specific heat versus the magnetic field, we can reject four other types of contributions: (i) a magnetic contribution, which would scale with $H/T$; (ii) a low-lying phonon mode, which would be insensitive to $H$; (iii) the normal-state electronic specific heat of an impurity, which would also be insensitive to $H$; and (iv) the broad superconducting transition of a spurious tin-rich phase, which would disappear at very low fields. Exotic scenarios such as a two-level specific heat due to the quasi degeneracy of the cubic- and tetragonal-state energies, together with a hypothetical stabilization at high fields associated with shifts in the electronic DOS on the meV scale, cannot be totally excluded. A definitive confirmation should come from the convergence of independent experimental results, both spectroscopic (tunneling, point-contact, etc.) and bulk (thermal conductivity, specific heat, thermal expansion, penetration depth). At the time of writing, we are just informed that features on an energy scale consistent with $\Delta_S(0)$ have indeed been detected by point-contact spectroscopy on the very same sample used in this work.[13] Further experiments, in particular thermal conductivity,[55] are in progress, and will be reported separately.

**Acknowledgements**

Stimulating discussions with J. Kortus, T. Jarlborg, and J. Geerk are gratefully acknowledged. We thank R. Cerny and E. Giannini for their expert support in X-ray diffraction, N. Clayton and A. Naula for their help. This work was supported by the National Science Foundation through the National Centre of Competence in Research "Materials with Novel Electronic Properties–MaNEP".

**Figure captions**

Fig. 1.  Observed, calculated, and difference X-ray powder diffraction patterns of $Nb_3Sn$ at 293 K. Inset: tetragonal lattice parameters $c$ and $a$ below the martensitic transformation temperature.

Fig. 2.  Total heat capacity of $Nb_3Sn$ versus temperature. (a), specific heat measured in zero field by continuous-heating adiabatic calorimetry above 15 K (~3000 independent data) and by relaxation calorimetry below 22 K. Inset: $ac$ susceptibility at the superconducting transition. (b), specific heat divided by the temperature near $T_c$ in fields of 0, 4, 10, and 16 T. The normal-state extrapolation (dashed line) is based on the entropy plot, Fig. 3. Inset: field-cooled (FC) and zero-field cooled (ZFC) susceptibility (corrected for a demagnetization factor $D = 0.05$; applied field: 5 mT).

Fig. 3.  Total entropy divided by the temperature for fields of (from right to left) 0, 2, 4, 7, 10, 13, and 16 T. The intercept of the extrapolation (full line) with the ordinate axis defines the Sommerfeld constant $\gamma$. A low-temperature anomaly is visible near the origin in zero field. Inset: low-temperature mixed-state coefficient $\gamma(H)$ evaluated at $T = 1$ K versus the field.

Fig. 4.  Electronic specific heat divided by the temperature in fields of (from right to left) 0, 2, 4, 7, 10, 13, and 16 T versus the temperature squared. Left inset: enlargement of the region below 9 K for fields of 0 and 10 T, showing the effect of the field on the anomaly. Right inset: enlargement of the region below 4.5 K for fields of (from bottom to top) 0, 0.2, 0.5, 1, 2, 4, 7, 10, 13, and 16 T.

Fig. 5.  Phase diagram in the $H - T$ plane. The present data (•) are derived from the middle of the specific heat jump. An alternative determination of $T_c(H)$ using the local entropy balance is shown as a vertical bar on the low temperature side. Khlopkin's data (×) are included for comparison.[19] The WHH curve fitted near $T_c$ is shown as a full line. Inset: deviation function of the thermodynamic critical field.

Fig. 6.  Normalized electronic specific heat as a function of the reduced temperature in zero field (frame (a), low temperature region; frame (b), near $T_c$), showing the measured

data ($\circ$), the BCS curve with $2\Delta_0/k_B T_c = 3.53$ (long-dashed line), a single-gap, strong-coupling curve with $2\Delta_0/k_B T_c = 4.7$ fitted to the specific heat jump (short-dashed line), and a two-gap fit (full line through the data). The residual deviation at the lowest temperatures might be due to the presence of smaller gaps, but other sources cannot be excluded. Inset of Fig. 6a: similar data for $MgB_2$.[43] Inset of Fig. 6b: semi-logarithmic plot of the normalized electronic specific heat versus the inverse reduced temperature, showing a crossover to a smaller asymptotic slope at low temperature (full line).

**Tables**

Table I.    Characteristic parameters of Nb$_3$Sn. $T_c$, superconducting transition temperature; $a$, lattice parameter at room temperature; $T_m$, martensitic transformation temperature; $a/c$, tetragonal distortion at 10 K; $V_{\text{gat}}$, mean atomic volume at $T$ = 10 K; $\gamma$, Sommerfeld constant; $\Theta_D(0)$, Debye temperature at $T \rightarrow 0$; $H_{c2}(0)$, upper critical field at $T \rightarrow 0$; $H_c(0)$, thermodynamic critical field at $T \rightarrow 0$; $H_{c1}(0)$, lower critical field at $T \rightarrow 0$ obtained from $H_c(0)\ln\kappa/(\sqrt{2}\kappa)$; $\xi$, Ginzburg-Landau coherence length; $\lambda$, Ginzburg-Landau penetration depth; $\kappa = \lambda/\xi$.

| | |
|---|---|
| $T_c$  (K) | 17.8±0.1 |
| $a$  (Å) | 5.293±0.001 |
| $T_m$  (K) | 40~45 |
| $a/c$ | 1.0026±0.0001 |
| $V_{\text{gat}}$  (cm$^3$ gat$^{-1}$) | 11.085±0.005 |
| $\gamma$  (mJ K$^{-2}$ gat$^{-1}$) | 13.7±0.2 |
| $\Theta_D(0)$  (K) | 233.7±2 |
| $H_{c2}(0)$  (T) | 25±0.5 |
| $H_c(0)$  (T) | 0.52±0.01 |
| $H_{c1}(0)$  (T) | 0.038 |
| $\xi$  (Å) | 36 |
| $\lambda$  (Å) | 1240 |
| $\kappa$ | 34 |

Table II.    Comparison of the parameters of the two-gap model in Nb$_3$Sn (this work) and MgB$_2$ (Ref. [2, 43]). $T_c$, superconducting transition temperature; $\gamma$, Sommerfeld constant; $\Delta_S(0)$, smaller gap at $T \rightarrow 0$; $\Delta_L(0)$, larger gap at $T \rightarrow 0$; $x$, fraction of the renormalized DOS of the band in which the smaller gap opens.

|  | Nb$_3$Sn | MgB$_2$ |
|---|---|---|
| $T_c$ (K) | 17.8 | 38 |
| $\gamma$ (mJ K$^{-2}$ gat$^{-1}$) | 13.7 | 0.9 |
| $2\Delta_S(0)/k_B T_c$ | 0.8±0.05 | 1.3 |
| $2\Delta_L(0)/k_B T_c$ | 4.9±0.05 | 3.9 |
| $x$ | 7.5% | 50% |
| $\Delta_S(0)$ (meV) | 0.61±0.05 | 2.1 |
| $\Delta_L(0)$ (meV) | 3.68±0.05 | 6.2 |

# References

[1]  A. Y. Liu, I. I. Mazin, and J. Kortus, Phys. Rev. Lett. **87**, 087005 (2001).

[2]  F. Bouquet, Y. Wang, I. Sheikin, P. Toulemonde, M. Eisterer, H. W. Weber, S. Lee, S. Tajima, and A. Junod, Physica C **385**, 192 (2003).

[3]  G. Binnig, A. Baratoff, H. E. Hoenig, and J. G. Bednorz, Phys. Rev. Lett. **45**, 1352 (1980).

[4]  L. Y. L. Shen, N. M. Senozan, and N. E. Phillips, Phys. Rev. Lett. **14**, 1025 (1965).

[5]  J. W. Hafstrom and M. L. A. MacVicar, Phys. Rev. B **2**, 4511 (1970).

[6]  H. Suhl, B. T. Matthias, and L. R. Walker, Phys. Rev. Lett. **3**, 552 (1959).

[7]  I. I. Mazin, O. K. Anderson, O. Jepsen, O. V. Dolgov, J. Kortus, A. A. Golubov, A. B. Kuz'menko, and D. van der Marel, Phys. Rev. Lett. **89**, 107002 (2002).

[8]  L. R. Testardi, Rev. Mod. Phys. **47**, 637 (1975).

[9]  M. Weger and I. B. Goldberg, in *Solid State Physics: Advances in Research and Applications*, edited by H. Ehrenreich, R. Seitz and D. Turnbull (Academic, New York, 1973), Vol. 28.

[10]  L. Y. L. Shen, Phys. Rev. Lett. **29**, 1082 (1972).

[11]  J. Kwo and T. H. Geballe, Physica **B109&110**, 1665 (1982).

[12]  F. Bouquet, Y. Wang, R. A. Fisher, D. G. Hinks, J. D. Jorgensen, A. Junod, and N. E. Phillips, Europhys. Lett. **56**, 856 (2001).

[13]  G. Goll, private communication (2004).

[14]  L. J. Vieland and A. W. Wicklund, Phys. Lett. **23**, 223 (1966).

[15]  A. Junod, J. Muller, H. Rietschel, and E. Schneider, J. Phys. Chem. Solids **39**, 317 (1978).

[16]  J. D. Axe and A. Shirane, Phys. Rev. B **8**, 1965 (1973).

[17]  G. R. Stewart, B. Cort, and G. W. Webb, Phys. Rev. B **24**, 3841 (1981).

[18]  G. R. Stewart and B. L. Brandt, Phys. Rev. B **29**, 3908 (1984).

[19]  M. N. Khlopkin, Sov. Phys. JETP **63**, 164 (1986).

[20]  W. Goldacker, R. Ahrens, M. Nindel, B. Obst, and C. Meingast, IEEE Trans. on Applied Superconductivity **3**, 1322 (1993).

[21]  J. R. Carvajal, *Program FullProf.2k, Version 2.20* (Laboratoire Léon Brillouin, CEA-CNRS, Saclay, France, 2002).

[22]  R. Mailfert, B. W. Batterman, and J. J. Hanak, Phys. Lett. A **24**, 315 (1967).

[23]  H. Devantay, J.-L. Jorda, M. Decroux, J. Muller, and R. Flükiger, J. Mater. Science **16**, 2145 (1981).

[24]  N. Toyota, T. Kobayashi, M. Kataoka, H. F. J. Watanabe, T. Fukase, Y. Muto, and F. Takei, J. Phys. Soc. Jap. **57**, 3089 (1988).

[25]  W. Weber and L. F. Mattheiss, Phys. Rev. B **25**, 2270 (1982).

[26]  B. Sadigh and V. Ozoliņš, Phys. Rev. B **57**, 2793 (1998).

[27]  D. W. Woodard and G. D. Cody, Phys. Rev. **136**, A166 (1964).

[28]  A. Junod, in *Studies of High Temperature Superconductors*, edited by A. Narlikar (Nova Science Publishers, Commack, N.Y., 1996), Vol. 19, p. 1.

[29]  G. Grimvall, *The Electron-Phonon Interaction in Metals* (North-Holland, Amsterdam, 1981).

[30]  J. K. Freericks, A. Y. Liu, A. Quandt, and J. Geerk, Phys. Rev. B **65**, 224510 (2002).

[31]  L. J. Vieland and A. W. Wicklund, Phys. Rev. **166**, 424 (1968).

[32]  D. U. Gubser, Phys. Rev. B **6**, 827 (1972).

[33]  J. M. Daams and J. P. Carbotte, Solid State Commun. **29**, 501 (1979).

[34]  B. Mühlschlegel, Z. Physik **155**, 313 (1959).

[35]  J. P. Carbotte, Rev. Mod. Phys. **62**, 1027 (1990).

[36]  H. Padamsee, J. E. Neighbor, and C. A. Shiffman, J. Low Temp. Phys. **12**, 387 (1973).

[37]  K. Maki, Phys. Rev. **139**, A702 (1965).

[38]  N. R. Werthamer, E. Helfand, and P. C. Hohenberg, Phys. Rev. **147**, 295 (1966).

[39]  S. V. Shulga, S.-L. Drechsler, H. Eschrig, H. Rosner, and W. Pickett., cond-mat/0103154.

[40]  S. V. Shulga, S. L. Drechsler, G. Fuchs, K. H. Müller, K. Winzer, M. Heinecker, and K. Krug, Phys. Rev. Lett. **80**, 1730 (1998).

[41]  C. Caroli and J. Matricon, Phys. kondens. Materie **3**, 380 (1965).

[42]  G. Otto, E. Saur, and H. Wizgall, J. Low Temp. Phys. **1**, 19 (1969).

[43]  A. Junod, Y. Wang, F. Bouquet, and P. Toulemonde, in *Studies of High Temperature Superconductors*, edited by A. V. Narlikar (Nova Science Publishers, Commack, N.Y., 2002), Vol. 38, p. 179; cond-mat/0106394.

[44]  J. M. Coombes and J. P. Carbotte, J. Low Temp. Phys. **74**, 491 (1989).

[45]  R. Combescot, Europhys. Lett. **43**, 701 (1998).

[46]  D. F. Moore, R. B. Zubeck, J. M. Rowell, and M. R. Beasley, Phys. Rev. B **20**, 2721 (1979).

[47]  D. A. Rudman and M. R. Beasley, Phys. Rev. B **30**, 2590 (1984).

[48]  J. Geerk, U. Kaufmann, W. Bangert, and H. Rietschel, Phys. Rev. B **33**, 1621 (1986).

[49]  M. R. Eskildsen, M. Kugler, S. Tanaka, J. Jun, S. M. Kazakov, J. Kapinski, and Ø. Fischer, Phys. Rev. Lett. **89**, 187003 (2002).

[50]  L. Hoffmann, A. K. Singh, H. Takei, and N. Toyota, J. Phys. F: Met. Phys. **18**, 2605 (1988).

[51]  L. F. Mattheiss and W. Weber, Phys. Rev. B **25**, 2248 (1982).

[52]  G. Gladstone, M. A. Jensen, and J. R. Schrieffer, in *Superconductivity*, edited by R. D. Parks (Marcel Dekker, New York, 1969), Vol. 2, p. 665; Transition Metals: Theory and Experiment.

[53]  J. C. F. Brock, Solid State Commun. **7**, 1789 (1969).

[54]  V. Hoffstein and R. W. Cohen, Phys. Lett. **29A**, 603 (1969).

[55]  A. V. Sologubenko, private communication (2004).

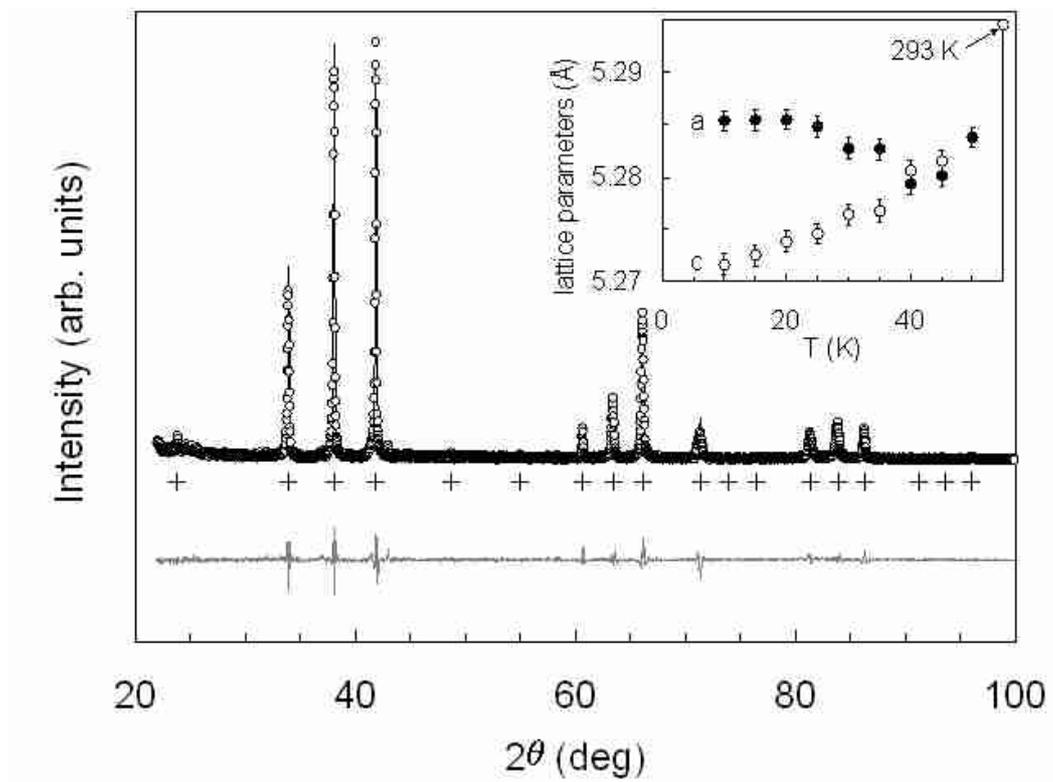

Fig. 1.   Observed, calculated, and difference X-ray powder diffraction patterns of $Nb_3Sn$ at 293 K. Inset: tetragonal lattice parameters $c$ and $a$ below the martensitic transformation temperature.

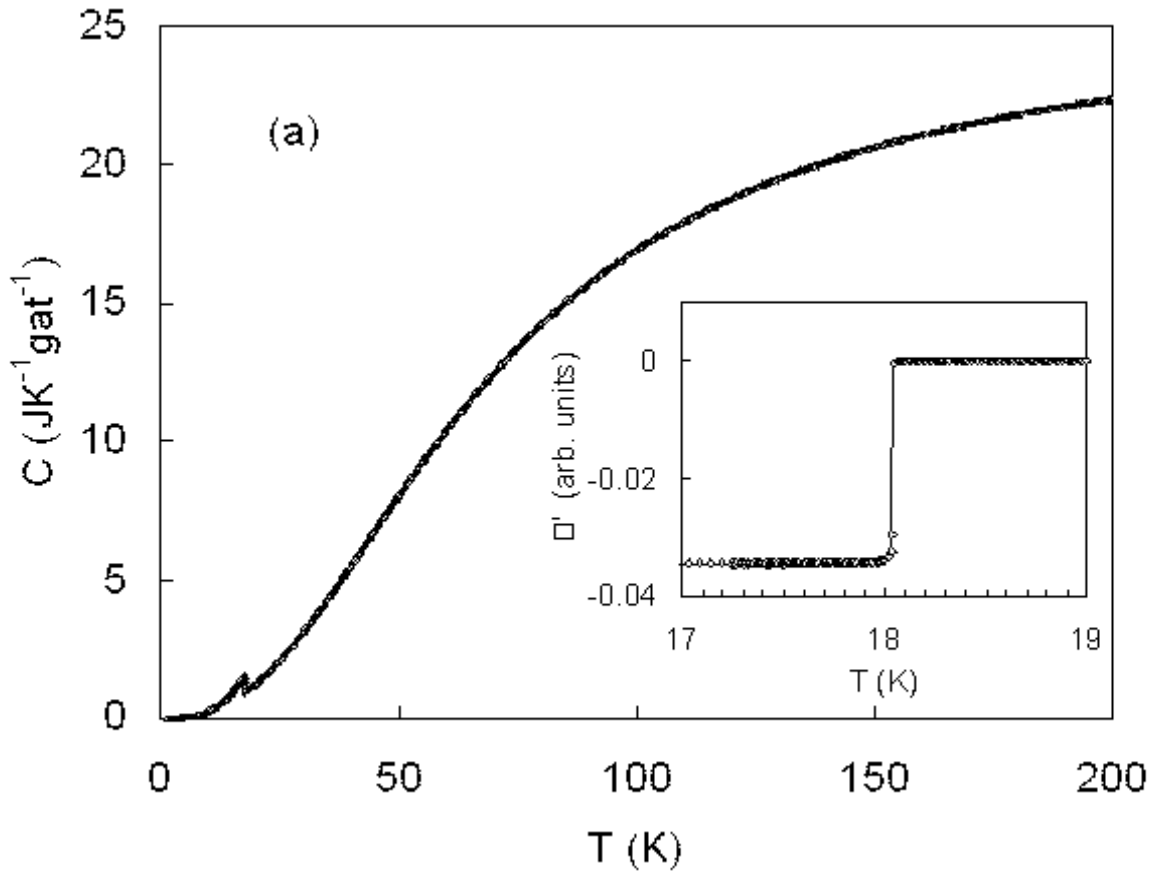

Fig. 2a

Fig. 2.  Total heat capacity of Nb$_3$Sn versus temperature. (a), specific heat measured in zero
field by continuous-heating adiabatic calorimetry above 15 K (~3000 independent
data) and by relaxation calorimetry below 22 K. Inset: *ac* susceptibility at the
superconducting transition. (b), specific heat divided by the temperature near $T_c$ in
fields of 0, 4, 10, and 16 T. The normal-state extrapolation (dashed line) is based on
the entropy plot, Fig. 3. Inset: field-cooled (FC) and zero-field cooled (ZFC)
susceptibility (corrected for a demagnetization factor $D = 0.03$; applied field: 5 mT).

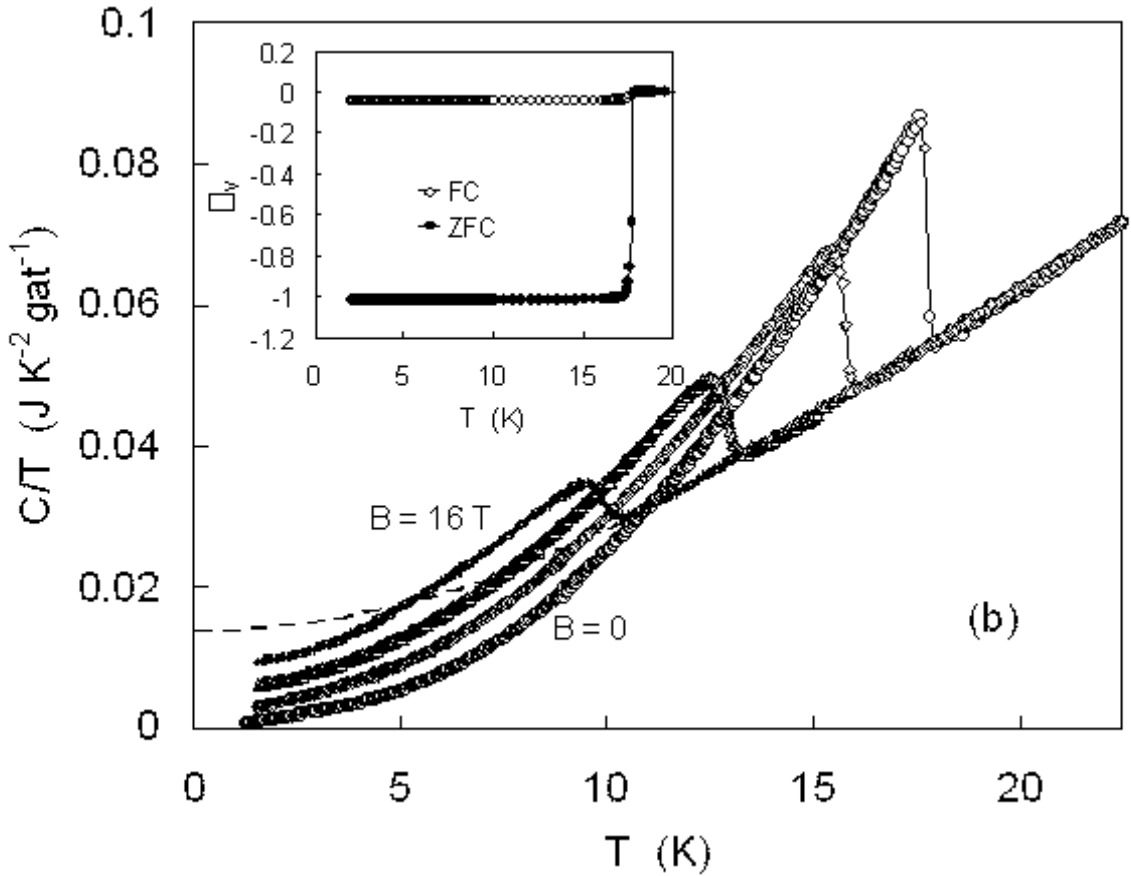

Fig. 2b

Fig. 2.   Total heat capacity of Nb$_3$Sn versus temperature. (a), specific heat measured in zero
field by continuous-heating adiabatic calorimetry above 15 K (~3000 independent
data) and by relaxation calorimetry below 22 K. Inset: *ac* susceptibility at the
superconducting transition. (b), specific heat divided by the temperature near $T_c$ in
fields of 0, 4, 10, and 16 T. The normal-state extrapolation (dashed line) is based on
the entropy plot, Fig. 3. Inset: field-cooled (FC) and zero-field cooled (ZFC)
susceptibility (corrected for a demagnetization factor $D = 0.05$; applied field: 5 mT).

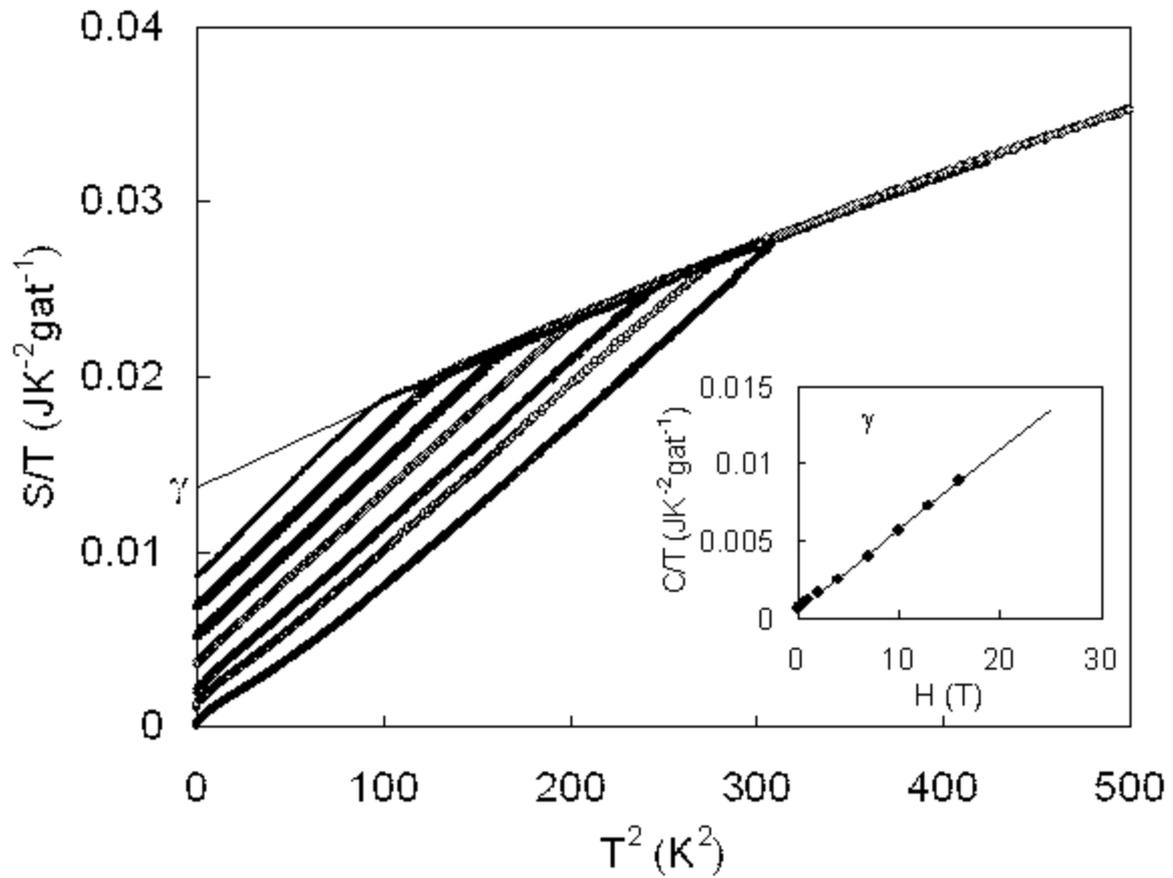

Fig. 3.   Total entropy divided by the temperature for fields of (from right to left) 0, 2, 4, 7, 10, 13, and 16 T. The intercept of the extrapolation (full line) with the ordinate axis defines the Sommerfeld constant $\gamma$. A low-temperature anomaly is visible near the origin in zero field. Inset: low-temperature mixed-state coefficient $\gamma(H)$ evaluated at $T = 1$ K versus the field.

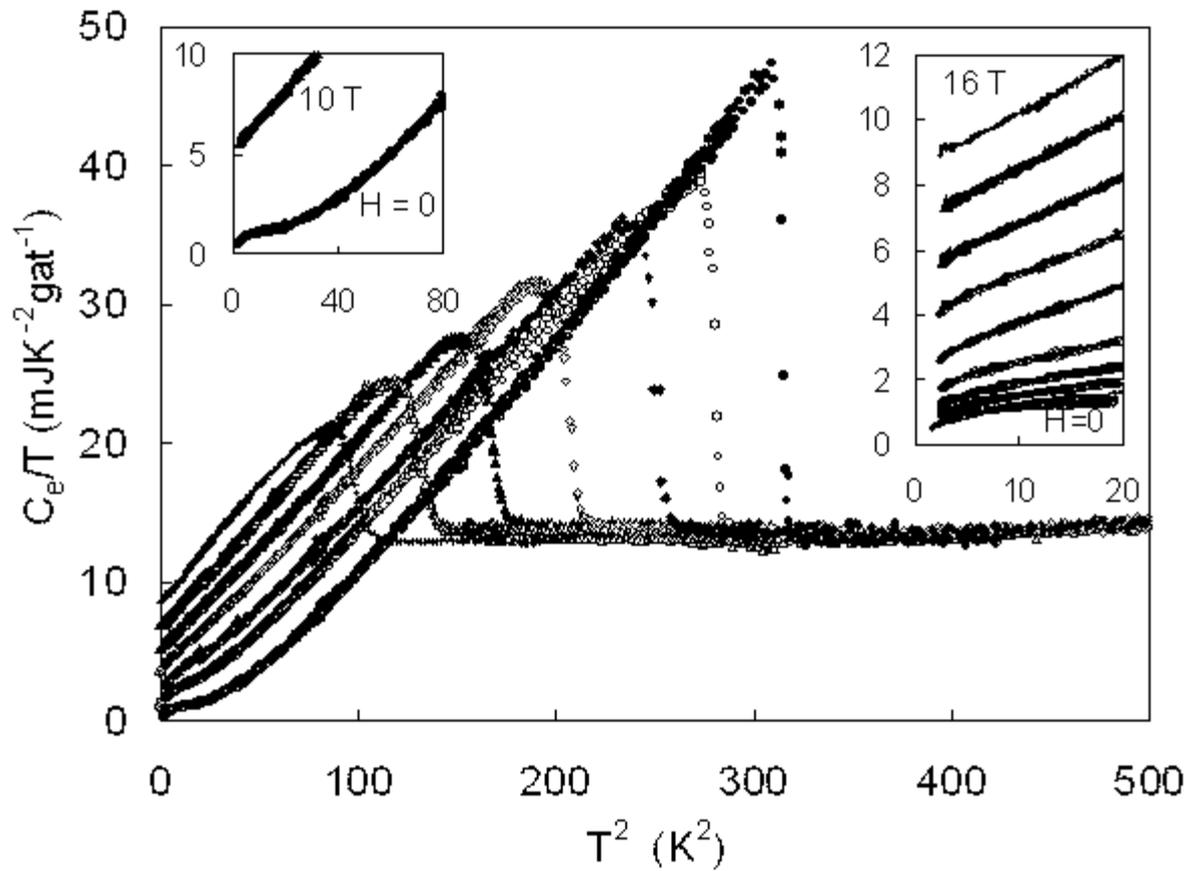

Fig. 4.   Electronic specific heat divided by the temperature in fields of (from right to left) 0, 2, 4, 7, 10, 13, and 16 T versus the temperature squared. Left inset: enlargement of the region below 9 K for fields of 0 and 10 T, showing the effect of the field on the anomaly. Right inset: enlargement of the region below 4.5 K for fields of (from bottom to top) 0, 0.2, 0.5, 1, 2, 4, 7, 10, 13, and 16 T.

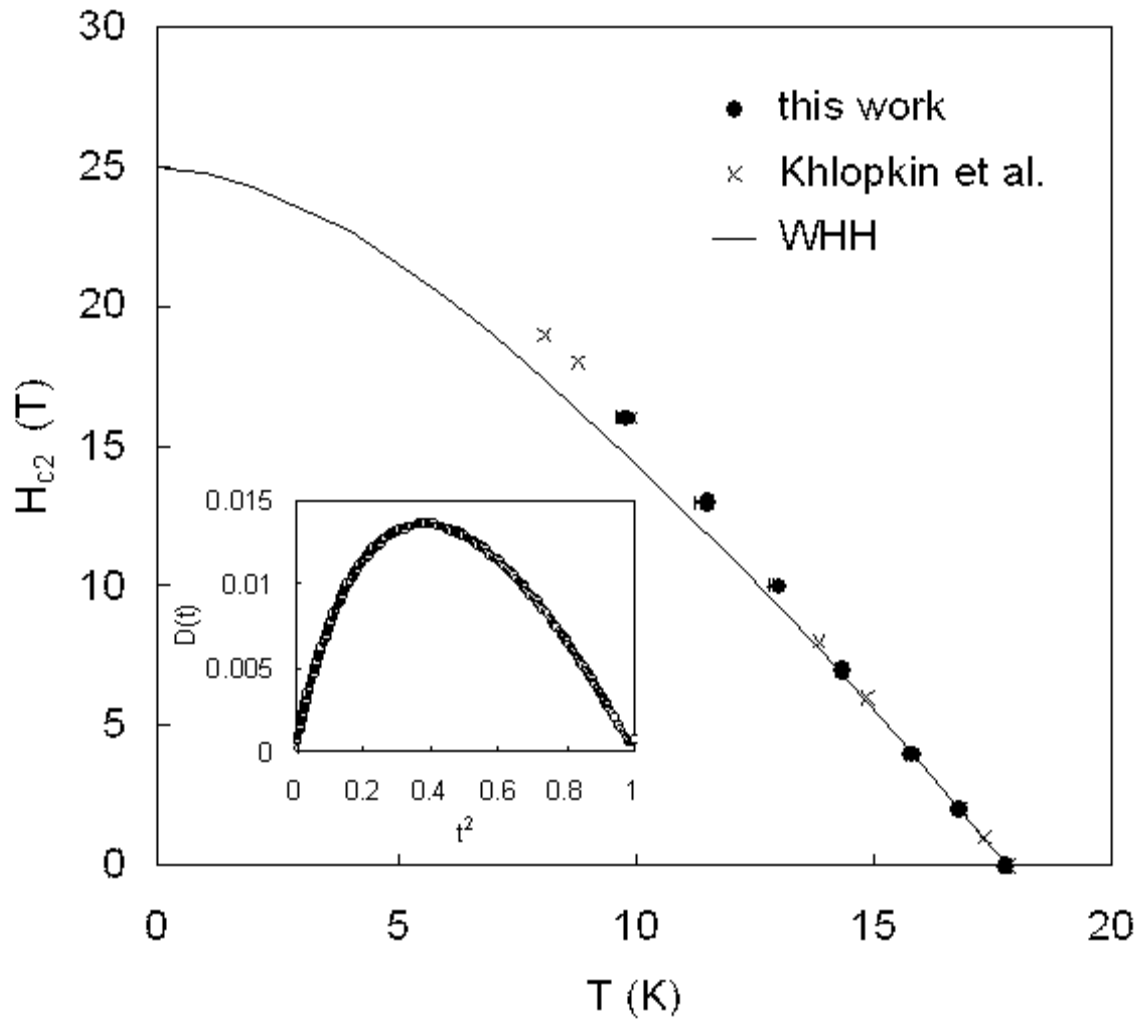

Fig. 5. Phase diagram in the $H - T$ plane. The present data (•) are derived from the middle of the specific heat jump. An alternative determination of $T_c(H)$ using the entropy balance is shown as a vertical bar on the low temperature side. Khlopkin's data (×) are included for comparison.[19] The WHH curve fitted near $T_c$ is shown as a full line. Inset: deviation function of the thermodynamic critical field.

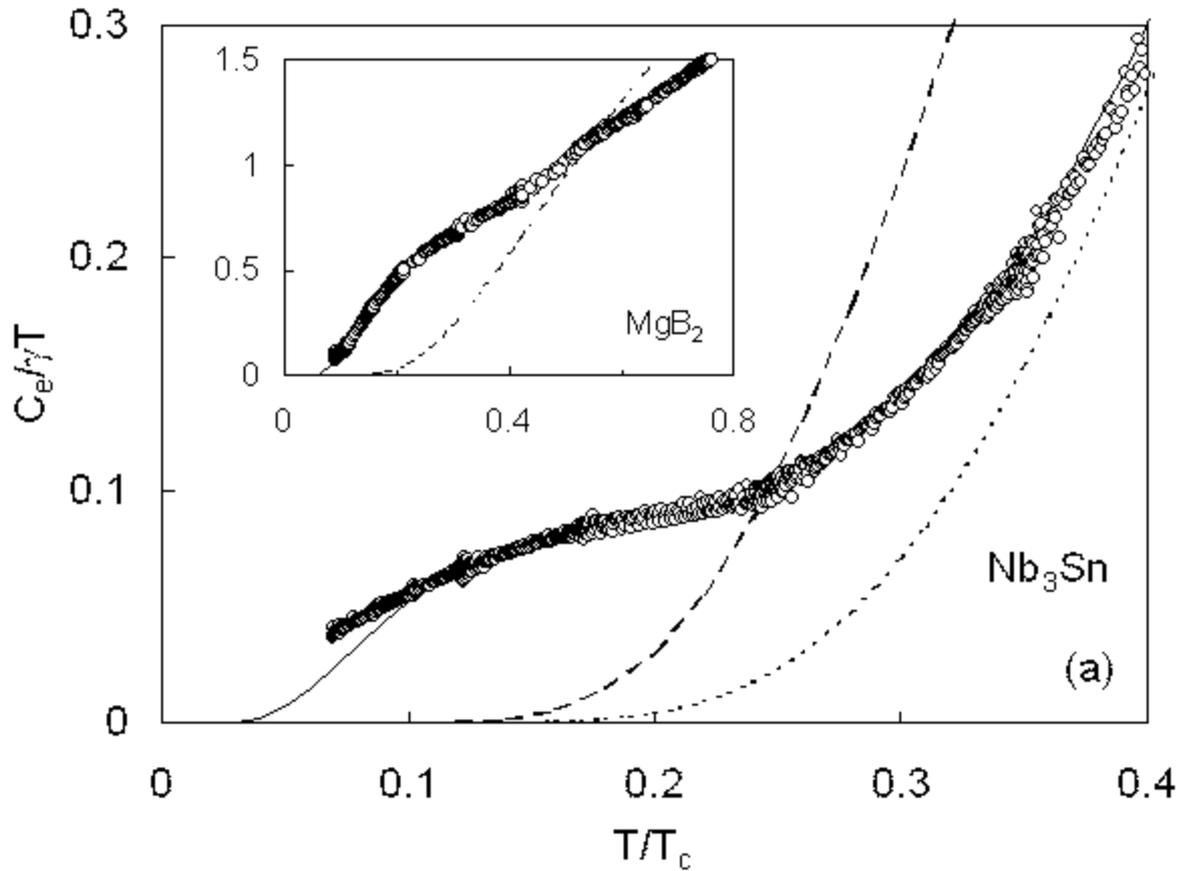

Fig. 6a

Fig. 6.   Normalized electronic specific heat as a function of the reduced temperature in zero field (frame (a), low temperature region; frame (b), near $T_c$), showing the measured data ($\circ$), the BCS curve with $2\Delta_0/k_B T_c = 3.53$ (long-dashed line), a single-gap, strong-coupling curve with $2\Delta_0/k_B T_c = 4.7$ fitted to the specific heat jump (short-dashed line), and a two-gap fit (full line through the data). The residual deviation at the lowest temperatures might be due to the presence of smaller gaps, but other sources cannot be excluded. Inset of Fig. 6a: similar data for $MgB_2$;[43] inset of Fig. 6b: semi-logarithmic plot of the normalized electronic specific heat versus the inverse reduced temperature, showing a crossover to a smaller asymptotic slope at low temperature (full line).

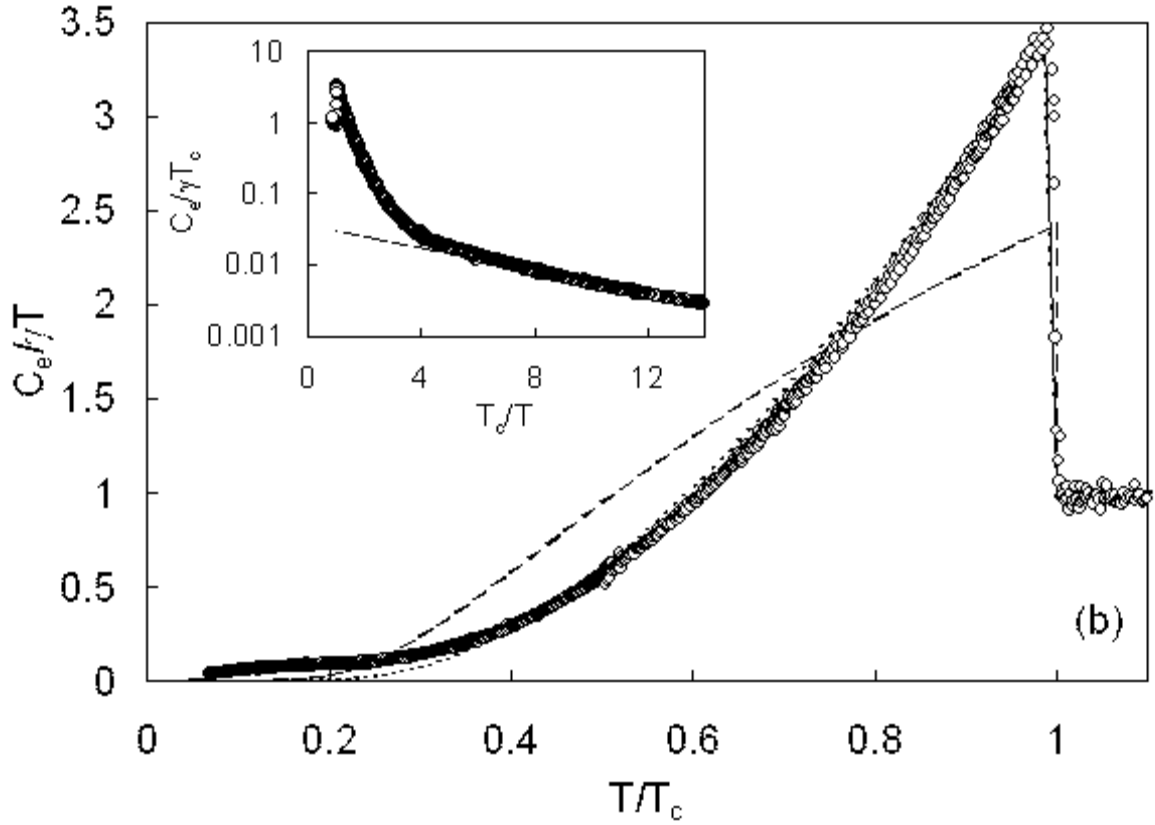

Fig. 6b

Fig. 6. Normalized electronic specific heat as a function of the reduced temperature in zero field (frame (a), low temperature region; frame (b), near $T_c$), showing the measured data (○), the BCS curve with $2\Delta_0/k_B T_c = 3.53$ (long-dashed line), a single-gap, strong-coupling curve with $2\Delta_0/k_B T_c = 4.7$ fitted to the specific heat jump (short-dashed line), and a two-gap fit (full line through the data). The residual deviation at the lowest temperatures might be due to the presence of smaller gaps, but other sources cannot be excluded. Inset of Fig. 6a: similar data for $MgB_2$;[43] inset of Fig. 6b: semi-logarithmic plot of the normalized electronic specific heat versus the inverse reduced temperature, showing a crossover to a smaller asymptotic slope at low temperature (full line).